\documentclass[onefignum,onetabnum]{siamonline220329}

\usepackage{amsfonts}
\usepackage{subfig}
\usepackage{amsmath}
\usepackage{amssymb}
\usepackage{booktabs}
\usepackage{hyperref}
\usepackage{tcolorbox}

\hypersetup{colorlinks,citecolor=blue}

\newcommand*\diff{\mathop{}\!\kern0pt\mathrm{d}}

\newcommand{\Var}{\mathrm{Var}}
\newcommand{\Error}{\mathrm{Error}}
\newcommand{\Black}{\mathrm{Black}}

\ifpdf
\hypersetup{
	pdftitle={Stochastic expansion for the pricing of Asian options},
	pdfauthor={Fabien Le Floc'h}
}
\fi
\title{Stochastic expansion for the pricing of Asian and basket options\thanks{Date: March 15, 2024.}}
\author{Fabien Le Floc'h \thanks{\email{fabien@2ipi.com}}}

\begin{document}
	\maketitle
	\begin{abstract}
We present closed analytical approximations for the pricing of basket options, also applicable to Asian options with discrete averaging under the Black-Scholes model with time-dependent parameters. The formulae are obtained by using a stochastic Taylor expansion around a log-normal proxy model and are found to be highly accurate for Asian options in practice as well as for vanilla options with discrete dividends.
	\end{abstract}
	\section{Introduction}
	
	The buyer of a vanilla basket option of strike $K$ and maturity $T$ on  underlying assets $(S_i)_{i=1,...,n}$  receives at maturity $T$ $\max\left(\sum_{i=1}^n w_i S_i(T) - K\right)$ for a call and $\max\left(K - \sum_{i=1}^n w_i S_i(T) \right)$ for a put. The weights $w_i$ typically sum to one. The buyer of an Asian call option of strike $K$ and averaging dates $(t_i)_{i=1,...,n}$  on an asset $S$ receives $\max\left(\sum_{i=1}^n w_i S(t_i) - K\right)$.  Under the Black-Scholes model with a term-structure of interest rates, dividend yields and volatilities, Asian options with discrete arithmetic averaging can be reduced to an equivalent vanilla basket option with a specific correlation structure. 
	Asian (basket) options are particularly popular among commodity traders, but is also traded regularly in the foreign exchange and in the equity derivatives spaces.
	
	There is no exact closed analytic formula for the pricing of arithmetic Asian or vanilla basket options. The problem has been widely studied. In the context of the Black-Scholes model, early valuation techniques were based on approximating the arithmetic mean by a geometric mean \cite{vorst1992prices}, or on moment matching against a lognormal distribution as in Levy \cite{levy1992pricing}. Later, Curran \cite{curran1994valuing} introduced a much more accurate pricing based on geometric conditioning which has been more recently further improved, by exploring better integration cutoffs and conditioning variables \cite{deelstra2010moment}. Ju proposed a third-order correction for the Levy moment matching in \cite{ju2002pricing}, while Zhou \cite{zhou2008accurate} looked at matching four moments with the log extended skew normal distribution. Beyond the Black-Scholes model, but restricted to Asian options, Zhang and Oosterlee propose an efficient method based on the Fourier cosine series for affine models with a known characteristic function in \cite{zhang2013efficient}. 
	
	Here, we consider a Black-Scholes setting with time-dependent parameters, and propose a stochastic expansion using the geometric average as a proxy, following the ideas of Etor\'e and Gobet, who applied their expansion to price vanilla options with discrete dividends under the piecewise-lognormal model in \cite{etore2012stochastic}, with further refinements in \cite{lefloch2015more}. We start by deriving the first, second and third-order expansions and then evaluate their accuracy on a sample of previously published cases. This allows for a comparison with other accurate approximations.

\section{Reductions to vanilla basket options}
\subsection{From Asian to basket}
Let $F(0,t)$ be the forward price of the underlying asset from valuation time to the date $t$. In the Black-Scholes model with time-dependent parameters we have \begin{equation*}F(0,t) = S(0)e^{\int_{0}^{t} r(s) - q(s) \diff s}\,,\end{equation*} with $r, q$ being the instantateneous interest rate and dividend yield and $S(t)$ the underlying asset spot at time $t$.  Under the risk-neutral measure $\mathbb{Q}$, the forward price follows
\begin{equation*}
	\diff F(t,T) = F(t,T) \sigma(t) \diff W(t)\,,
\end{equation*}
where $W$ is a $\mathbb{Q}$-Brownian motion.

The buyer of an arithmetic Asian option of strike $K$ and maturity $T$ on an underlying $S$ with observation dates $(t_i)_{i=1,...,n}$ with $t_n=T$, receives at maturity $\max\left(\sum_{i=1}^n w_i S(t_i) - K\right)$ for a call and $\max\left(K - \sum_{i=1}^n w_i S(t_i) \right)$ for a put. It may be expressed as a basket option of strike $K$ and maturity $T$ on $n$ underlying assets $(S_i)_{i=1,...,n}$. The covariance between the assets must match the covariance of $S$ between different observation times, this means\footnote{We use the It\^o isometry to obtain Equations \ref{eqn:correlation_asian} and \ref{eqn:correlation_asian_basket}.}
\begin{equation}
	\rho_{i,j} \sqrt{v_i v_j} =v_{i \wedge j}\,, \label{eqn:correlation_asian}
\end{equation}
where $\rho_{i,j}$ denotes the correlation between the assets $S_i$ and $S_j$, $v_i = \int_{0}^{t_i} \sigma^2 (s) \diff s$ being the total variance up to time $t_i$ for the asset $S$. 
Each underlying asset forward to $T$ must match the forwards of the single asset to $t_i$, meaning 
\begin{equation}F_i(0,T) = \mathbb{E}[S_i(T)] = \mathbb{E}[S(t_i)] = F(0,t_i)  = S(0)e^{\int_0^{t_i} (r(s)-q(s)) \diff s}\,.\end{equation}

Thus, the arithmetic Asian option on $S$ with observations $t_i$ corresponds to a vanilla basket option on $S_i$ with forwards to maturity $F_i(0,T)$, total variances $v_i$ and correlations $\rho_{i,j}$ for $i,j=1,...,n$.

\subsection{Asian baskets}
This may be extended to Asian baskets with payoff at maturity \begin{equation*}\max\left(\sum_{i=1}^{n} w_i \sum_{j=1}^{m}  \mu_j S_j(t_i)- K,0\right)\,,\end{equation*} where $(\mu_j)_{j=1...m}$ are basket weights and $S_j$ are the different assets with correlation $\rho_{j,l}$. We may define $n\times m$ assets $\hat{S}_{i+n(j-1)}$ with forward $F_j(t_i)$, total variance $\int_{0}^{t_i} \sigma_j^2(s) \diff s$, correlation
\begin{equation}\hat{\rho}_{i+n(j-1),k+n(l-1)} = \rho_{j,l} \frac{\int_0^{t_{i} \wedge t_k} \sigma_j(s) \sigma_l(s)\diff s}{\sqrt{ \int_0^{t_i} \sigma_j^2(s) \diff s \int_0^{t_k} \sigma_l^2(s) \diff s}}\,, \label{eqn:correlation_asian_basket}\end{equation}
for $(i,k) \in \{1,...,n\}^2$, $(j,l) \in\left\{1,...,m\right\}^2$. The Asian basket option is equivalent to a vanilla basket option on $\hat{S}_{i+n(j-1)}$ with weights $w_i \mu_j$.

\subsection{Average strike options}
A floating strike Asian (call) option paying at maturity $T$, $\max\left(S(T)- k \sum_{i=1}^n S(t_i)\right)$, with $k$ being a strike percentage, can be expressed as an equivalent fixed strike Asian (put) option \cite{vecer2002unified}. Through this symmetry, the formulae given in this paper may be used to price floating (also known as average) strike options.

\section{Expansions}

\subsection{The proxy}
An early technique, first explored by Vorst \cite{vorst1992prices} to estimate the price of Asian options with arithmetic averaging is to approximate it by the geometric average option equivalent, because the geometric average of geometric Brownian motions is another geometric Brownian motion and can thus be priced with the usual Black-Scholes formula. This was extended to the more general cases of vanilla basket options by Gentle \cite{gentle1993basket}.

Under the risk-neutral measure $\mathbb{Q}$, the forward price for each asset follows
\begin{equation*}
	\diff F_i(t,T) = F_i(t,T) \sigma_i(t) \diff W_i(t)\,,
\end{equation*}
where $W_i$ is a $\mathbb{Q}$-Brownian motion, and we have $\diff W_i  \diff W_j = \rho_{i,j} \diff t$.
The payoff at maturity of the basket  option reads
\begin{align*}
	V(T) &= \left[ \eta\left(\sum_{i=1}^n w_i S_i(T) - K\right) \right]^+\\
	&= \left(\sum_{i=1}^n w_i F_i(0,T) \right) \left[ \eta\left(\sum_{i=1}^n \tilde{a}_i S_i^\star(T) - {K}^\star\right) \right]^+
		= A \left[ \eta\left(\sum_{i=1}^n \tilde{a}_i S_i^\star(T) - {K}^\star\right) \right]^+\\		
\end{align*} with $\eta = \pm 1$ for a call (respectively a put) option and
\begin{align*}
	\tilde{a}_i = \frac{w_i F_i(0,T)}{A(T)}\,,\quad A = \sum_{i=1}^n w_i F_i(0,T) \,,\quad {K}^\star = \frac{K}{A}\,,\nonumber\\
	S_i^\star(t) = \frac{S_i(t)}{F_i(0,t)} = e^{-\frac{1}{2}\int_{0}^{t} \sigma_i^2(s) \diff s + \int_{0}^{t} \sigma_i(s) \diff W_i(s)}\,.
\end{align*} 
We approximate the arithmetic average by the geometric average leading to
\begin{equation}
	V(0) \approx  B(0,T) A \mathbb{E}\left[ \left( \eta\left(\prod_{i=1}^n  S_i^\star(T)^{a_i} - {K}^\star\right) \right)^+\\		   \right]
\end{equation}
with $a_i = \tilde{a}_i$ as a specific case.

Let \begin{equation*}G(T) = \prod_{i=1}^n  S_i^\star(T)^{a_i}  = e^{ \sum_{i=1}^n -\frac{1}{2} a_i \int_{0}^{T} \sigma_i^2(s) \diff s + \int_{0}^{T} a_i \sigma_i(s) \diff W_i(s)}\end{equation*}
We have \begin{align}
	m(T) = \mathbb{E}\left[\ln G(T)\right] &=  -\frac{1}{2} \sum_{i=1}^n a_i v_i  \,,\label{eqn:ElogG}\\
	\tilde{\nu}^2(T) = \Var\left[\ln G(T)\right] &= \sum_{i,j=1}^n  a_i a_j \rho_{i,j}\int_0^T \sigma_i(s) \sigma_j(s) \diff s \label{eqn:VarG}\,.
	\end{align}
	with $v_i = \int_{0}^T \sigma_i^2(s) \diff s$.
	
Vorst \cite{vorst1992prices}, Gentle \cite{gentle1993basket} propose to adjust the strike price of the geometric average option such that the expectation of the term inside the max function is preserved and use $\tilde{K}^\star = K^\star - (1 - \mathbb{E}\left[G(T) \right])$ instead of $K^\star$. 
A different way to preserve the expectation consists in adjusting the spot side instead of the strike side: 
\begin{equation}
	V(0) \approx  B(0,T) A \mathbb{E}\left[ \left(  \eta\left(\alpha\prod_{i=1}^n  S_i^\star(T)^{a_i} - {K}^\star\right) \right)^+ \right]
\end{equation}
with $\alpha = \frac{1}{\mathbb{E}[G(T)]}$ and $B(0,T)$ is the discount factor to the payment date. The latter is found to be more accurate in Borovykh \cite{borovykh2013pricing}.

Levy \cite{levy1992pricing} matches not only the mean but also the variance of the arithmetic average by a lognormal distribution. We have
\begin{equation}
	\nu_A^2 = \ln\left( \Var\left[A(T)\right] \right)= \ln\left( \sum_{i,j=1}^n a_i a_j e^{\rho_{i,j}\int_0^T \sigma_i(s) \sigma_j(s) \diff s}\right)\,..
\end{equation}
We may adjust the $a_i$ used in the geometric mean such that the variance of the geometric mean matches exactly the one of the arithmetic mean by using \begin{equation}a_i = \tilde{a}_i \frac{\nu_A}{\tilde{\nu}}\,.\label{eqn:ai_levy}\end{equation}

Through Equation \ref{eqn:VarG} and the definition of the factor $\alpha$, we obtain the following formula for our Vorst Geometric (VG) and Vorst Levy (VL) approximations, using respectively $a_i = \tilde{a}_i$ and $a_i$ as defined by Equation \ref{eqn:ai_levy}:
\begin{eqnarray}
	V(0) \approx  A  \cdot \Black(1, K^\star, {\nu}^2, T)
\end{eqnarray}
where $\Black(F,K,v,T)$ denotes the Black-76 formula applied to the forward $F$, strike $K$ and total variance $v$ to maturity $T$  and $\nu = \tilde{\nu}$ for the geometric approximation, $\nu = \nu_A$ for the Levy approximation. We have
\begin{equation}\Black(F,K,v,T)=\eta B(0,T) \left[F\Phi(\eta d_1)-K\Phi(\eta d_2)\right]\end{equation} with $d_1 = \frac{1}{\sqrt{v}}\left(\ln\frac{F}{K} + \frac{1}{2}v\right)$, $d_2 = d_1 -  \sqrt{v}$.

We will adopt the Vorst geometric formula with adjusted forward as proxy for our stochastic Taylor expansion. The Vorst geometric  formula with strike shift would lead to slightly more complicated equations and exhibit a worse accuracy.

The Vorst Levy formula corresponds exactly to the Levy approximation, but defines a stochastic process based on the individual underlying asset processes. This will allow to define the  stochastic expansions based on this alternative proxy as well.

\subsection{First order expansion}
Let $h(x) = \max(\eta x, 0)$, we apply a Taylor expansion of order-1 on  $h$:
\begin{multline}
	\mathbb{E}\left[ B(0,T) h\left(\sum_{i=1}^n \tilde{a}_i S_{i}^{\star}(T) -K^\star \right) \right] = \mathbb{E}\left[B(0,T) h\left(\alpha \prod_{i=1}^n S_{i}^{\star}(T)^{a_i} - K^\star\right) \right] \\
	+ 	\mathbb{E}\left[ B(0,T)\left( \sum_{i=1}^n \tilde{a}_i S_{i}^{\star}(T) - \alpha \prod_{i=1}^n S_{i}^{\star}(T)^{a_i}\right) h'\left(\alpha \prod_{i=1}^n S_{i}^{\star}(T)^{a_i} - K^\star\right)\right]\\ + \Error_2(h)
\end{multline}
where $\Error_2(h) \leq_c 2 \left(||  \sum_{i=1}^n \tilde{a}_i S_{i}^{\star}(T) - \alpha \prod_{i=1}^n S_{i}^{\star}(T)^{a_i} ||_3\right)^2$. 
We may rewrite the second term using a derivative with regards to $K^\star$, and interchange derivation and expectation (see Etor\'e and Gobet \cite{etore2012stochastic})  to obtain
\begin{multline}	\mathbb{E}\left[ \left( \sum_{i=1}^n \tilde{a}_i S_{i}^{\star}(T) - \alpha \prod_{i=1}^n S_{i}^{\star}(T)^{a_i}\right) h'\left(\alpha \prod_{i=1}^n S_{i}^{\star}(T)^{a_i} - K^\star\right)\right]\\  = -\frac{\partial}{\partial K^\star} \mathbb{E}\left[ \left( \sum_{i=1}^n \tilde{a}_i S_{i}^{\star}(T) - \alpha \prod_{i=1}^n S_{i}^{\star}(T)^{a_i}\right) h\left(\alpha \prod_{i=1}^n S_{i}^{\star}(T)^{a_i} - K^\star\right)\right] \label{eqn:first_order_correction_raw}
	\end{multline}
Let $G^\star = \alpha G$, we have  $\mathbb{E}\left[ G^\star(T)\right] = 1$ and $G^\star$ may be thus expressed as \begin{equation*} G^\star(T)=e^{-\frac{1}{2}\int_0^T\beta(s)^2\diff s + \int_0^T \beta(s)  dW_G(T)}\end{equation*} in probability, with $W_G$ being a Brownian motion and $\beta^2(s)=\frac{\partial \nu^2(s)}{\partial s}$. We interpret $\frac{G^\star(T)}{G^\star(0)}$ as a change of measure on $\mathcal{F}_T$. Under the new induced measure $\mathbb{Q}^G$, $\tilde{W}_G (T)= W_G(T) - \int_0^T \beta(s) \diff s $ is a Brownian motion. Then $G^\star(T)$ under $\mathbb{Q}^G$ has the same law as $G^\star(T) e^{\nu^2 (T)}$ under $\mathbb{Q}$. Thus,
\begin{align}	\mathbb{E}\left[  G^\star(T) h'\left( G^\star(T)- K^\star\right)\right] &= -\frac{\partial}{\partial K^\star} \mathbb{E}^G\left[h\left( \tilde{G}^\star(T)e^{\nu^2 (T)} - K^\star\right) \right]\,.\label{eqn:EGG}
\end{align}
The right hand side corresponds to the derivative of a vanilla option price under the Black model with  forward $e^{\nu^2(T)}$, strike $K^\star$ and total variance $\nu^2(T)$.

In similar fashion we use a change of measure $\frac{S_{i}^{\star}(T)}{S_i^\star(0)}$ and define \begin{equation*}\tilde{W}_i(t) = W_i(t) - \int_0^t \sigma_i(s) \diff s \end{equation*}  to obtain
\begin{align}	\mathbb{E}\left[  S_{i}^{\star}(T) h'\left( G^\star(T)- K^\star\right)\right] &= -\frac{\partial}{\partial K^\star} \mathbb{E}^i\left[h\left( \tilde{G}^\star(T)e^{ \sum_{j=1}^n a_j\int_0^{T}\rho_{i,j}\sigma_i(s)\sigma_j(s)\diff s} - K^\star\right) \right]\,.\label{eqn:ESG}
\end{align}
We use Equation \ref{eqn:EGG} and Equation \ref{eqn:ESG} in Equation \ref{eqn:first_order_correction_raw}  to obtain the first order expansion
\begin{multline}
	V_\textsf{VG1} = A \cdot \Black(1,K^\star,\nu^2,T) \\
	+ A\frac{\partial}{\partial K^\star}\Black\left(e^{\nu^2},K^\star,\nu^2,T\right)
	 - A\sum_{i=1}^n \tilde{a}_i \frac{\partial}{\partial K^\star}\Black\left(e^{\bar{v}_i},K^\star,\nu^2,T\right) \,.
\end{multline}
with \begin{align}\bar{v}_i = \sum_{l=1}^n a_l v_{i,l}\,,&\quad v_{i,l} = \rho_{i,l}\int_{0}^T \sigma_i(s) \sigma_l(s)\diff s\,.\label{eqn:vibar_vil}\end{align} In particular, for an Asian, we have  $v_{i,l} = \int_{0}^{t_i \wedge t_j} \sigma^2(s)\diff s$. For baskets, the volatilities are taken to be constant equal to each underlying implied volatility to maturity in practice and we have  $v_{i,l} = \rho_{i,l} \sqrt{v_i v_l}$.

The relevant derivatives of the Black-76 formula are given in Appendix \ref{sec:appendix_black_der}. The formula with the Vorst Levy proxy is exactly the same, but uses different $\nu$ and $a_i$.
\subsection{Second order expansion}
The second-order Taylor expansion involves the additional term
\begin{align*}
	\frac{1}{2}\frac{\partial}{\partial K^\star} \mathbb{E}\left[ \left(\sum_{i=1}^n \tilde{a}_i S_{i}^{\star}(T) - \alpha \prod_{i=1}^n S_{i}^{\star}(T)^{a_i}\right)^2 h\left(\alpha \prod_{i=1}^n S_{i}^{\star}(T)^{a_i} - K^\star\right)\right]
\end{align*}
We decompose it into three terms. The first term involves only $G^\star$, we perform two successive changes of measure around $\frac{G^\star(T)}{G^\star(0)}$  to obtain
\begin{align*}
	\mathbb{E}\left[G^\star(T)^2 h(G^\star(T)-K)\right] &= \mathbb{E}^{G}\left[\tilde{G}^\star e^{\nu^2(T)} h\left(\tilde{G}^\star e^{\nu^2(T)}-K^\star\right)\right]\\
	&= e^{\nu^2(T)} \mathbb{E}^{G^2}\left[h\left(\tilde{\tilde{G}}^\star e^{2\nu^2(T)}-K^\star\right)\right]\,.
\end{align*}

The second term involves the cross product $S_{i}^{\star}(T)G^\star(T)$. We perform two successive changes of measure around $\frac{G^\star(T)}{G^\star(0)}$ and $\frac{S_{i}^{\star}(T)}{S_i^\star(0)}$ to obtain
\begin{multline*}
	\mathbb{E}\left[S_{i}^{\star}(T) G^\star(T) h(G^\star(T)-K)\right] = \\\mathbb{E}^{i}\left[\tilde{G}^\star e^{\sum_{j=1}^n a_j \int_0^{T} \rho_{i,j}\sigma_i(s) \sigma_j(s) \diff s} h\left(\tilde{G}^\star e^{\sum_{j=1}^n a_j \int_0^{T} \rho_{i,j} \sigma_i(s)\sigma_j(s) \diff s} -K^\star\right)\right]=\\
  e^{\sum_{j=1}^n a_j \int_0^{T} \rho_{i,j} \sigma_i(s)\sigma_j(s) \diff s}  \mathbb{E}^{i,G}\left[h\left(\tilde{\tilde{G}}^\star e^{\nu^2(T)+\sum_{j=1}^n a_j \int_0^{T}\rho_{i,j} \sigma_i(s)\sigma_j(s)  \diff s}-K^\star\right)\right]\,.
\end{multline*}

The third term involves $S_{i}^{\star}(T) S_j^\star(T)$. We perform two successive changes of measure around $\frac{S_{i}^{\star}(T)}{S_i^\star(0)}$ and $\frac{S_j^\star(T)}{S_j^\star(0)}$ to obtain
\begin{multline*}
	\mathbb{E}\left[S_{i}^{\star}(T) S_j^\star(T) h(G^\star(T)-K)\right] \\= \mathbb{E}^{i}\left[\tilde{S}_j^\star(T) e^{ \int_0^{T} \rho_{i,j} \sigma_i(s)\sigma_j(s) \diff s} h\left(\tilde{G}^\star e^{\sum_{l=1}^n a_l \int_0^{T} \rho_{i,l} \sigma_i(s)\sigma_l(s) \diff s} -K^\star\right)\right]\\
 =  e^{\int_0^{T} \rho_{i,j} \sigma_i(s)\sigma_j(s) \diff s}  \mathbb{E}^{i,j}\left[h\left(\tilde{\tilde{G}}^\star e^{ \sum_{l=1}^n a_l \left(\int_0^{T} \rho_{i,l} \sigma_i(s)\sigma_l(s) \diff s + \int_0^{T} \rho_{j,l} \sigma_j(s)\sigma_l(s) \diff s\right)}-K^\star\right)\right]\,.
\end{multline*}

Our formula for the second order expansion thus reads
\begin{multline}
	V_\textsf{VG2} = A \cdot \Black(1,K^\star,\nu^2,T) \\
	+ A\frac{\partial}{\partial K^\star}\Black\left(e^{\nu^2},K^\star,\nu^2,T\right)
	- A\sum_{i=1}^n \tilde{a}_i \frac{\partial}{\partial K^\star}\Black\left(e^{\bar{v}_i},K^\star,\nu^2,T\right) \\
	+\frac{A}{2} e^{\nu^2} \frac{\partial^2}{\partial {K^{\star}}^2}\Black\left(e^{2\nu^2},K^\star,\nu^2,T\right)
	+\frac{A}{2} \sum_{i=1}^n \tilde{a}_i e^{\bar{v}_i}   \frac{\partial^2}{\partial {K^\star}^2}\Black\left(e^{\nu^2+\bar{v}_i},K^\star,\nu^2,T\right)\\
	-A \sum_{i,j=1}^n \tilde{a}_i \tilde{a}_j  e^{v_{i,j} }  \frac{\partial^2}{\partial {K^\star}^2}\Black\left(e^{\bar{v}_i+ \bar{v}_j},K^\star,\nu^2,T\right)\,.
\end{multline}
\subsection{Third order expansion}
The third order expansion adds the term
\begin{align*}
-	\frac{1}{6}\frac{\partial}{\partial K^\star} \mathbb{E}\left[ \left(\sum_{i=1}^n \tilde{a}_i S_{i}^{\star}(T) - \alpha \prod_{i=1}^n S_{i}^{\star}(T)^{a_i}\right)^3 h\left(\alpha \prod_{i=1}^n S_{i}^{\star}(T)^{a_i} - K^\star\right)\right]
\end{align*}
In similar fashion as for the second order, we perform three consecutive changes of measure involving $\frac{G^\star(0)}{G^\star(0)}$, $\frac{S_{i}^{\star}(T)}{S^\star(0)}$. This leads to our third order expansion
Our formula for the second order expansion thus reads
\newtcolorbox{mymathbox}[1][]{colback=white, sharp corners, #1}

\begin{mymathbox}
\begin{align}
	V_\textsf{VG3} &= A \cdot \Black(1,K^\star,\nu^2,T) \\
	&+ A\frac{\partial}{\partial K^\star}\Black\left(e^{\nu^2},K^\star,\nu^2,T\right)
	- A\sum_{i=1}^n \tilde{a}_i \frac{\partial}{\partial K^\star}\Black\left(e^{\bar{v}_i},K^\star,\nu^2,T\right) \nonumber\\
	&+\frac{A}{2} e^{\nu^2} \frac{\partial^2}{\partial {K^{\star}}^2}\Black\left(e^{2\nu^2},K^\star,\nu^2,T\right)\nonumber\\
	&- A \sum_{i=1}^n \tilde{a}_i e^{\bar{v}_i}   \frac{\partial^2}{\partial {K^\star}^2}\Black\left(e^{\nu^2+\bar{v}_i},K^\star,\nu^2,T\right)\nonumber\\
	&+\frac{A}{2} \sum_{i,j=1}^n  \tilde{a}_i \tilde{a}_j  e^{v_{i,j}}  \frac{\partial^2}{\partial {K^\star}^2}\Black\left(e^{\bar{v}_i + \bar{v}_j},K^\star,\nu^2,T\right)\nonumber\\
	&+\frac{A}{6} e^{2\nu^2} \frac{\partial^3}{\partial {K^{\star}}^3}\Black\left(e^{3\nu^2},K^\star,\nu^2,T\right)\nonumber\\
	&-\frac{A}{2} \sum_{i=1}^n \tilde{a}_i e^{\nu^2+2 \bar{v}_i}   \frac{\partial^3}{\partial {K^\star}^3}\Black\left(e^{2\nu^2+\bar{v}_i},K^\star,\nu^2,T\right)\nonumber\\
	&+\frac{A}{2} \sum_{i,j=1}^n \tilde{a}_i \tilde{a}_j e^{ \bar{v}_i+\bar{v}_j+v_{i,j}}   \frac{\partial^3}{\partial {K^\star}^3}\Black\left(e^{\nu^2+\bar{v}_i+v_j},K^\star,\nu^2,T\right)\nonumber\\
	&-\frac{A}{6} \sum_{i,j,l=1}^n \tilde{a}_i \tilde{a}_j \tilde{a}_l e^{v_{i,l}+v_{j,l}+v_{i,j}} 	 \frac{\partial^3}{\partial {K^\star}^3}\Black\left(e^{\bar{v}_i+\bar{v}_j+\bar{v}_l},K^\star,\nu^2,T\right)\nonumber\,,
\end{align}
\end{mymathbox}
with $\bar{v}_i$, $v_{i,j}$ defined in Equation \ref{eqn:vibar_vil}.
\subsection{Symmetries}
We may exploit the symmetries in the double and triple sums to reduce the number terms being summed and speed up the corresponding algorithm.
In particular,
\begin{multline*}
 \sum_{i,j=1}^n  \tilde{a}_i \tilde{a}_j  e^{v_{i,j}}  \frac{\partial^2}{\partial {K^\star}^2}\Black\left(e^{\bar{v}_i + \bar{v}_j},K^\star,\nu^2,T\right) = \sum_{i=1}^n  \tilde{a}_i^2  e^{v_{i,i}}  \frac{\partial^2}{\partial {K^\star}^2}\Black\left(e^{2\bar{v}_i },K^\star,\nu^2,T\right)\\ + 2\sum_{i=1}^n \sum_{j=1}^{i-1}   \tilde{a}_i \tilde{a}_j  e^{v_{i,j}}  \frac{\partial^2}{\partial {K^\star}^2}\Black\left(e^{\bar{v}_i + \bar{v}_j},K^\star,\nu^2,T\right)\,.
\end{multline*}
and
\begin{multline*}
 \sum_{i,j,l=1}^n \tilde{a}_i \tilde{a}_j \tilde{a}_l e^{v_{i,l}+v_{j,l}+v_{i,j}} 	 \frac{\partial^3}{\partial {K^\star}^3}\Black\left(e^{\bar{v}_i+\bar{v}_j+\bar{v}_l},K^\star,\nu^2,T\right) =\\
6 \sum_{i=1}^{n} \sum_{j=1}^{i-1} \sum_{l=1}^{j-1} \tilde{a}_i \tilde{a}_j \tilde{a}_l e^{v_{i,l}+v_{j,l}+v_{i,j}} 	 \frac{\partial^3}{\partial {K^\star}^3}\Black\left(e^{\bar{v}_i+\bar{v}_j+\bar{v}_l},K^\star,\nu^2,T\right)\\
+ 3 \sum_{i=1}^n \sum_{j=1}^{i-1} \tilde{a}_i \tilde{a}_j^2  e^{2v_{i,j}+v_{j,j}} 	 \frac{\partial^3}{\partial {K^\star}^3}\Black\left(e^{\bar{v}_i+2\bar{v}_j},K^\star,\nu^2,T\right)\\
+ 3 \sum_{i=1}^n \sum_{l=1}^{i-1} \tilde{a}_i^2 \tilde{a}_l e^{2v_{i,l}+v_{i,i}} 	 \frac{\partial^3}{\partial {K^\star}^3}\Black\left(e^{2\bar{v}_i+\bar{v}_l},K^\star,\nu^2,T\right)\\
+ \sum_{i=1}^n \tilde{a}_i^3 e^{3v_{i,i}} 	 \frac{\partial^3}{\partial {K^\star}^3}\Black\left(e^{3\bar{v}_i},K^\star,\nu^2,T\right)\,.
\end{multline*}

\section{Numerical examples for Asian options}
\subsection{Weekly}
We start by reproducing the example of Ju \cite[Table 4]{ju2002pricing} of a weekly averaging call option of maturity 3 years with different strikes and volatilities in Table \ref{tbl:jutable4}.
\begin{table}[h]
	\caption{Prices of weekly averaging call options of maturity 3 years with different strikes $K$ and volatilities $\sigma$, with a constant interest rate $r=9\%$.\label{tbl:jutable4}}
	\centering{
		\begin{tabular}{@{}l r r r r r r r r@{}}\toprule
			$(\sigma, K)$ & MC & Ju & LB & Deelstra & VG1 & VG2 & VG3 & VL3 \\ \midrule
	(0.05, 95) & 15.1197 & 15.1197 & 15.1197 & 15.1197 &  15.1197 & 15.1197 & 15.1197 & 15.1197\\
	(0.05, 100) & 11.3069 & 11.3069 & 11.3069 & 11.3069 &  11.3069 & 11.3070 & 11.3069 & 11.3069\\
	(0.05, 105) & 7.5561 & 7.5562 & 7.5561 & 7.5561 &  7.5561 & 7.5561 & 7.5561 & 7.5561\\
	(0.10, 95) & 15.2163 & 15.2165 & 15.2161 & 15.2163 &  15.2159 & 15.2163 & 15.2163 & 15.2163\\
	(0.10, 100) & 11.6390 & 11.6394 & 11.6388 & 11.6390 &  11.6387 & 11.6390 & 11.6390 & 11.6390\\
	(0.10, 105) & 8.3911 & 8.3913 & 8.3909 & 8.3911 &  8.3908 & 8.3911 & 8.3911 & 8.3911\\
	(0.20, 95) & 16.6342 & 16.6365 & 16.6322 & 16.6342 &  16.6317 & 16.6341 & 16.6342 & 16.6342\\
	(0.20, 100) & 13.7626 & 13.7634 & 13.7607 & 13.7626 &  13.7600 & 13.7625 & 13.7626 & 13.7626\\
	(0.20, 105) & 11.2146 & 11.2134 & 11.2128 & 11.2146 &  11.2118 & 11.2145 & 11.2146 & 11.2146\\
	(0.30, 95) & 19.0145 & 19.0179 & 19.0089 & 19.0145 &  19.0058 & 19.0140 & 19.0144 & 19.0144\\
	(0.30, 100) & 16.5766 & 16.5755 & 16.5712 & 16.5766 &  16.5675 & 16.5762 & 16.5766 & 16.5766\\
	(0.30, 105) & 14.3830 & 14.3774 & 14.3775 & 14.3830 &  14.3733 & 14.3827 & 14.3830 & 14.3830\\
	(0.40, 95) & 21.7269 & 21.7307 & 21.7148 & 21.7269 &  21.7056 & 21.7256 & 21.7268 & 21.7267\\
	(0.40, 100) & 19.5738 & 19.5690 & 19.5614 & 19.5738 &  19.5516 & 19.5727 & 19.5737 & 19.5737\\
	(0.40, 105) & 17.6110 & 17.5978 & 17.5979 & 17.6109 &  17.5878 & 17.6100 & 17.6109 & 17.6108\\
	(0.50, 95) & 24.5527 & 24.5583 & 24.5296 & 24.5527 &  24.5106 & 24.5498 & 24.5524 & 24.5523\\
	(0.50, 100) & 22.6115 & 22.6032 & 22.5871 & 22.6116 &  22.5679 & 22.6090 & 22.6113 & 22.6111\\
	(0.50, 105) & 20.8241 & 20.8023 & 20.7980 & 20.8242 &  20.7791 & 20.8219 & 20.8239 & 20.8238\\
			\midrule
			RMSE & & 0.0068 & 0.0115 & 0.0000 & 0.0204 & 0.0011 & 0.0001 & 0.0002 \\
			MAE & & 0.0218 & 0.0262 & 0.0001 & 0.0450 & 0.0029 & 0.0002 & 0.0004 \\
			\bottomrule
	\end{tabular}}
\end{table}
It consists in 157 averaging dates, starting at the valuation date $t=0$ and ending at the maturity date $t=T=3$ years. We add the prices for the lower bound and refined conditioned moment matching lognormal approximation of \cite{deelstra2010moment} (respectively named "LB" and "Deelstra") as well as the first, second and third order stochastic expansions named "VLE-1", "VLE-2", "VLE-3". The reference prices are obtained by a quasi Monte-Carlo simulation with Brownian-bridge variance reduction on 32 million paths.

The Deelstra lower bound is found to be more accurate than the Curran geometric conditioning in \cite{ju2002pricing}. We also report slightly lower errors for the Ju approximation than what is presented in \cite{ju2002pricing}, because the reference prices used in \cite{ju2002pricing} are less precise than ours. The Ju approximation is slightly more accurate than the Deelstra lower bound. The second order stochastic expansion is significantly more accurate, and the third order expansion is of similar accuracy as the more refined Deelstra approximation.
The VL proxy leads to slightly larger errors than the VG proxy.

\subsection{Monthly}
\begin{figure}[h]
	\centering{\includegraphics[width=\textwidth]{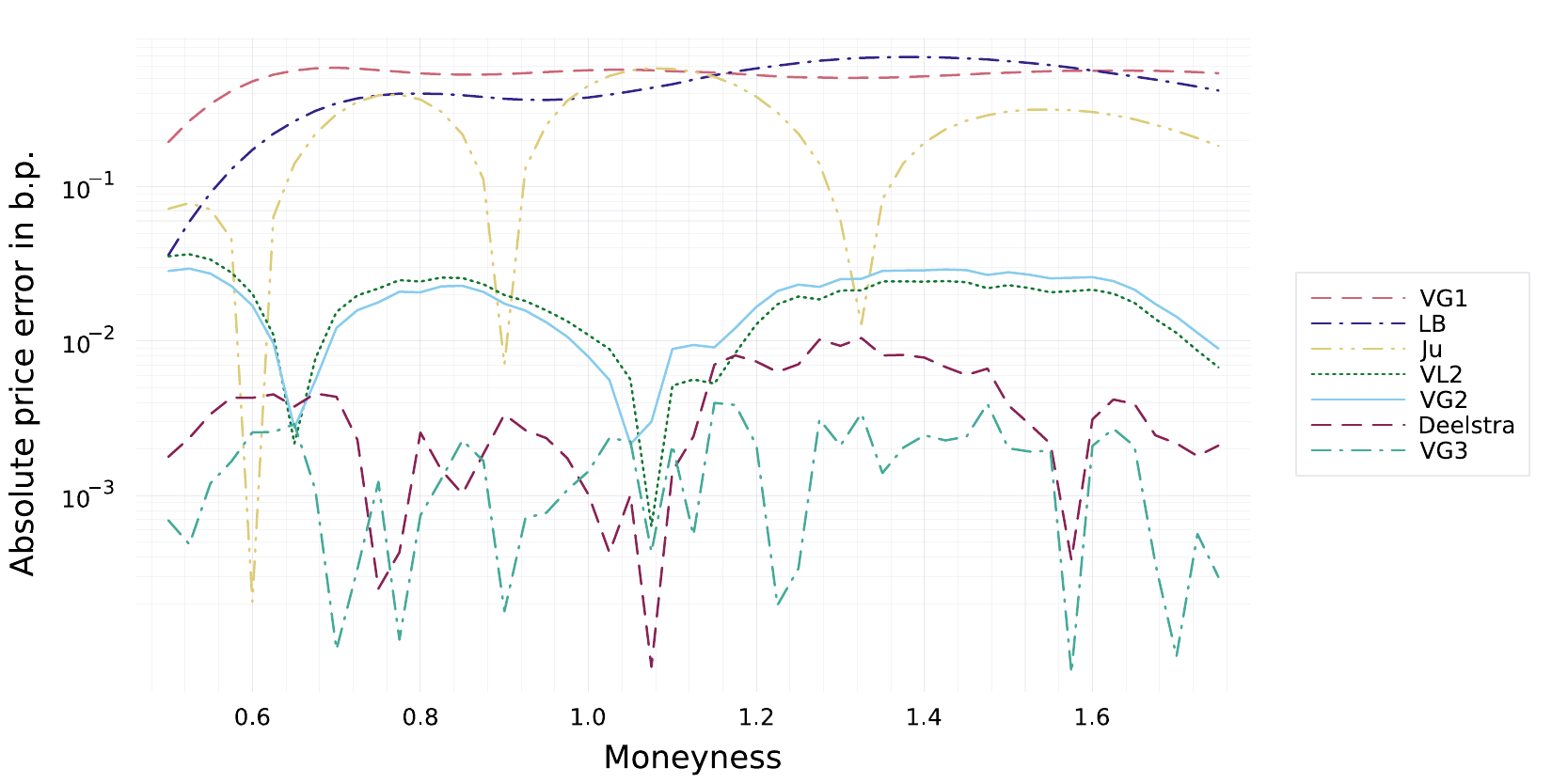}}
	\caption{Error in the price of Asian options of maturity 1 year on HAL with monthly averaging, for a range of spot moneyness.}
	\label{fig:zhou_hal_monthly_1y}
\end{figure}

We then consider the example of an Asian option of maturity 1 year with monthly averaging, on the HAL stock ticker, presented in \cite{zhou2008accurate}. The interest rate is chosen to be $r=6\%$, the dividend yield   $q=0.97\%$, the volatility is $\sigma=41.33\%$ and the spot price is $S=30.78$. The 12 averaging dates are evenly\footnote{We reproduce here the example of Zhou as is. It is a manufactured example albeit on realistic market data, as, in practice, the averaging dates are typically not exactly evenly spread out due to holidays and week-ends.} spread out. Figure \ref{fig:zhou_hal_monthly_1y} shows the error in price expressed in b.p.  of the underlying spot. The reference prices are obtained with a quasi Monte-Carlo simulation on 128 million paths. 

The first order expansion is of similar accuracy as the Deelstra lower bound approximation.
The second order expansion is already very accurate, with maximum error well below 0.1 b.p., much lower than the error of the Deelstra lower bound approximation. In Zhou \cite{zhou2008accurate}, their four moments matching  is found to be comparable but slightly less accurate than the Deelstra lower bound. The third order and the Deelstra approximations are extremely accurate, below the Monte-Carlo sampling error, as evidenced by the wiggles in the error profile, with however a slightly larger error for the Deelstra approximation. Asian options with weekly averaging lead to similar observations.
Again, the VL proxy leads to slightly larger errors than the VG proxy.

\subsection{Yearly}
\begin{figure}[h!]
	\centering{
		\subfloat[][First and second orders.]{\includegraphics[width=0.48\textwidth]{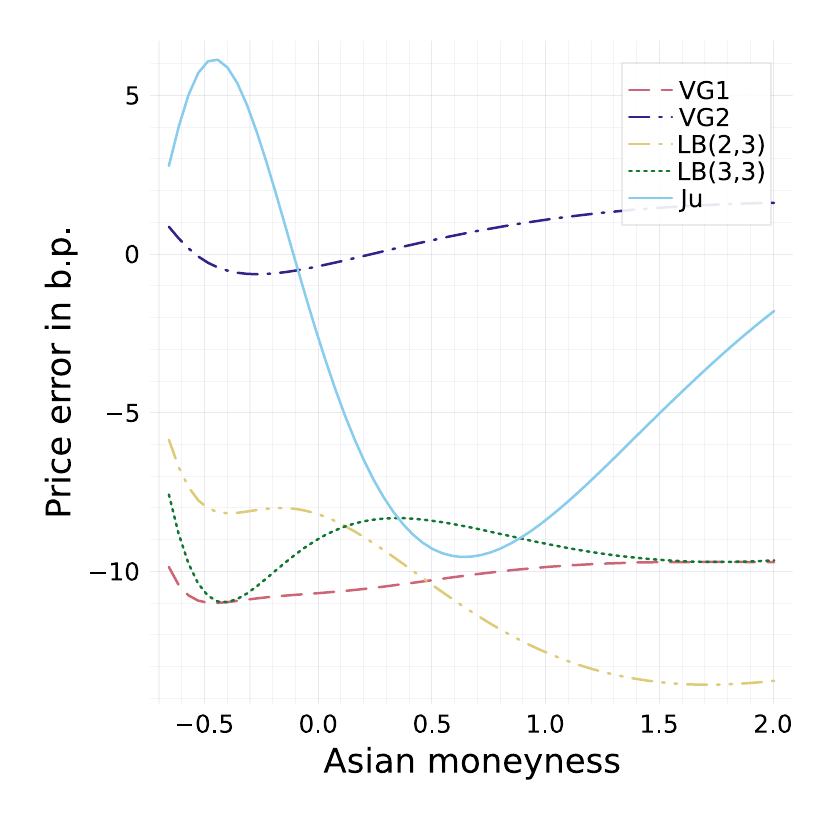}}
		\subfloat[][Second and third orders.]{\includegraphics[width=0.48\textwidth]{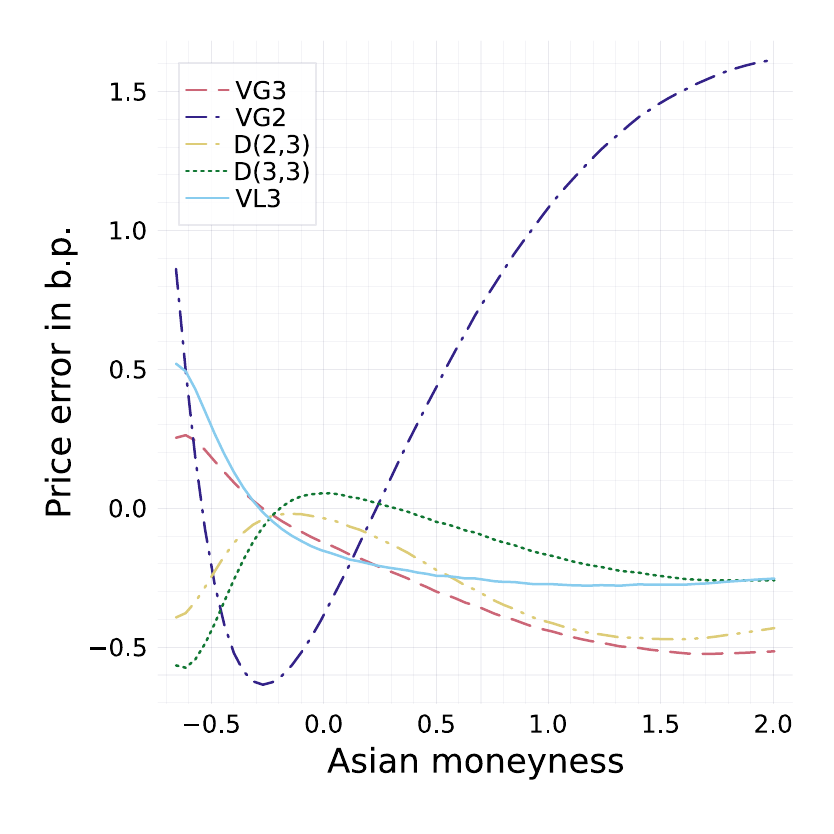}}
	}
	\caption{Error in the price of Asian options of maturity 5 years with yearly averaging.}
	\label{fig:lord_yearly_5y_second}
\end{figure}

Finally, we study the more extreme examples of Lord \cite{lord2005partially}: yearly averaging call options of maturity 5 and 30 years.
The 5 year options are on an asset of spot price $S=100$, with interest rate $r=5\%$, no dividend and and a somewhat large volatility $\sigma=50\%$ while the 30 year option use a volatility $\sigma=25\%$. They respectively consist in 5 and 30 averaging dates. Tables \ref{tbl:lordtable4_5} and \ref{tbl:lordtable8_30} reproduce \cite[Tables 7 and 9]{lord2005partially} and reports the error in basis points of the underlying spot for three strikes  $K$ of Asian moneyness \begin{equation*}
	M=\frac{K}{\sum w_i F(0,t_i)}-1\,.
\end{equation*} We name LB(2,3) and Deelstra(2,3) the lower bound and conditioned lognormal moment matching approximations in Deelstra et al. \cite{deelstra2010moment} using $\delta_2$ and $f_3$ as defined in their paper. We also test the Deelstra(3,3) variant with $\delta_3$ and $f_3$.
\begin{table}[ht]
	\caption{Error in b.p. in the prices of yearly averaging call options of maturity 5 years with different strikes. D($i$,$j$) corresponds to the Deelstra($i$,$j$) approximation with $\delta_i$ and $f_j$ \cite{deelstra2010moment}.\label{tbl:lordtable4_5}}
	\centering{
		\begin{tabular}{@{}r r r r r r r r r r@{}}\toprule
			Strike & $M$& Ju & LB(2,3) & D(2,3) & D(3,3) & VG1 & VG2 & VG3 & VL3\\ \midrule
58.2370 &-0.5& 5.99 & -7.93 & -0.24 & -0.43 & -10.96 & -0.20 & 0.19 & 0.31 \\
116.4741 & 0.0& -2.62 & -8.19 & -0.03 & 0.06 & -10.67 & -0.37 & -0.11 & -0.14 \\
174.7111 & 0.5&-9.27 & -10.41 & -0.21 & -0.04 & -10.27 & 0.45 & -0.29 & -0.23 \\
			\bottomrule
\end{tabular}}
\end{table}
\begin{table}[ht]
	\caption{Error in b.p. in the prices of yearly averaging call options of maturity 30 years with different strikes. D($i$,$j$) corresponds to the Deelstra($i$,$j$) approximation with $\delta_i$ and $f_j$ \cite{deelstra2010moment}.\label{tbl:lordtable8_30}}
	\centering{
		\begin{tabular}{@{}r r r r r r r r r r @{}}\toprule
			Strike &  $M$ & Ju & LB(2,3) & D(2,3) & D(3,3) & VG1 & VG2 & VG3 & VL3\\ \midrule
118.9819 & -0.5&10.30 & -3.62 & 0.05 & -0.10 & -7.59 & -0.51 & 0.06 & 0.12 \\
237.9638 & 0.0& -2.43 & -14.04 & -0.29 & 0.05 & -7.95 & -0.43 & -0.04 & -0.08 \\
356.9457 & 0.5& -12.39 & -22.87 & -0.48 & 0.03 & -7.84 & 0.02 & -0.10 & -0.11 \\
			\bottomrule
\end{tabular}}
\end{table}
For the $(2,3)$ variant, the errors for LB and Deelstra match respectively what is reported in \cite{lord2005partially} as "LB\_GA" and "Curran2M+". The maximum error of the $(3,3)$ variant is slightly larger. The second and third order expansions are found to be nearly as accurate as the Deelstra conditioned lognormal moment matching approximation. The third order here does not appear to improve significantly over the second order. Figure \ref{fig:lord_yearly_5y_second} shows that the third-order is markedly more accurate if we widen the moneyness range. Again, the first order expansion exhibits a similar error as the Deelstra lower bound approximation.

\begin{figure}[h!]
	\centering{
		\subfloat[][First and second orders.]{\includegraphics[width=0.48\textwidth]{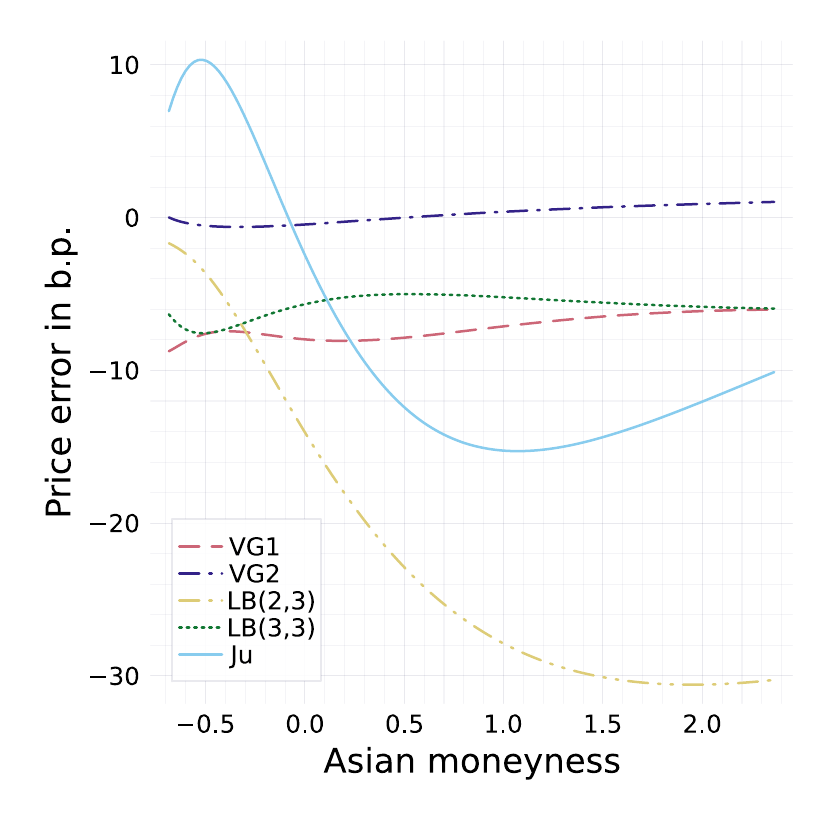}}
		\subfloat[][Second and third orders.]{\includegraphics[width=0.48\textwidth]{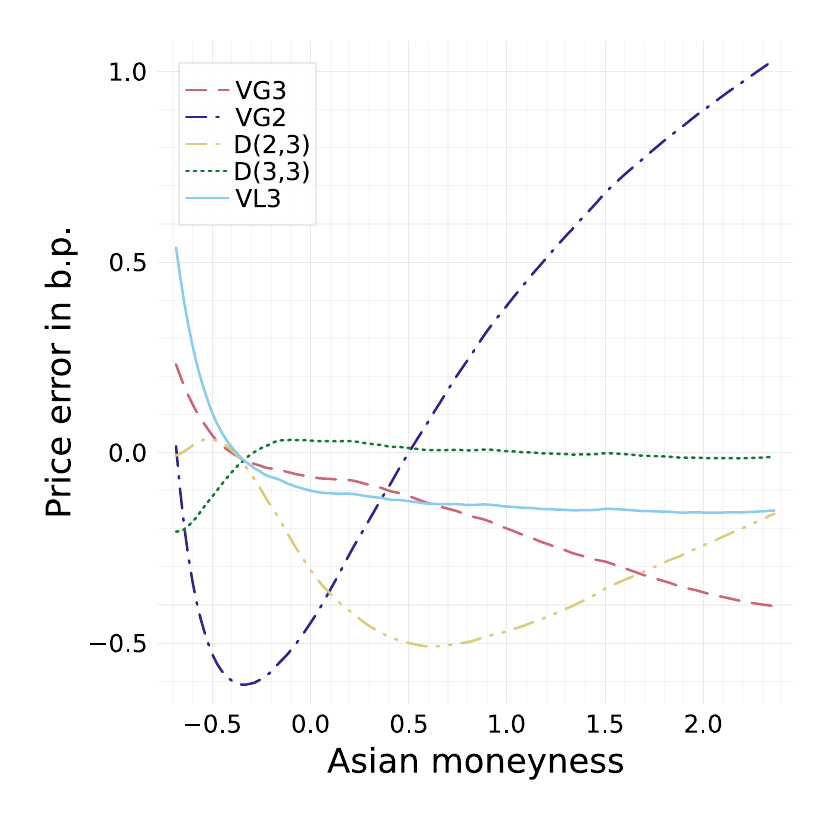}}
	}
	\caption{Error in the price of Asian options of maturity 30 years with yearly averaging.}
	\label{fig:lord_yearly_30y_second}
\end{figure}

For yearly options of maturity 30 years, Table \ref{tbl:lordtable8_30} shows the first order expansion to be more accurate than the Ju and the LB(2,3) lower bound approximations. The Deelstra(3,3) performs slightly better than the Deelstra(2,3) on this example. The second order expansion is as accurate as the Deelstra(2,3) approximation within the range of Asian moneyness [-0.5, 0.5] and becomes slightly less accurate for larger Asian moneyness (Figure \ref{fig:lord_yearly_30y_second}). The third order expansions VG3 or VL3 allow to further reduce the error, but are still less accurate than the best Deelstra(3,3) approximation further away from the money.

\subsection{Two remaining observations}
\begin{figure}[h!]
	\centering{
		\subfloat[][First and second orders.]{\includegraphics[width=0.48\textwidth]{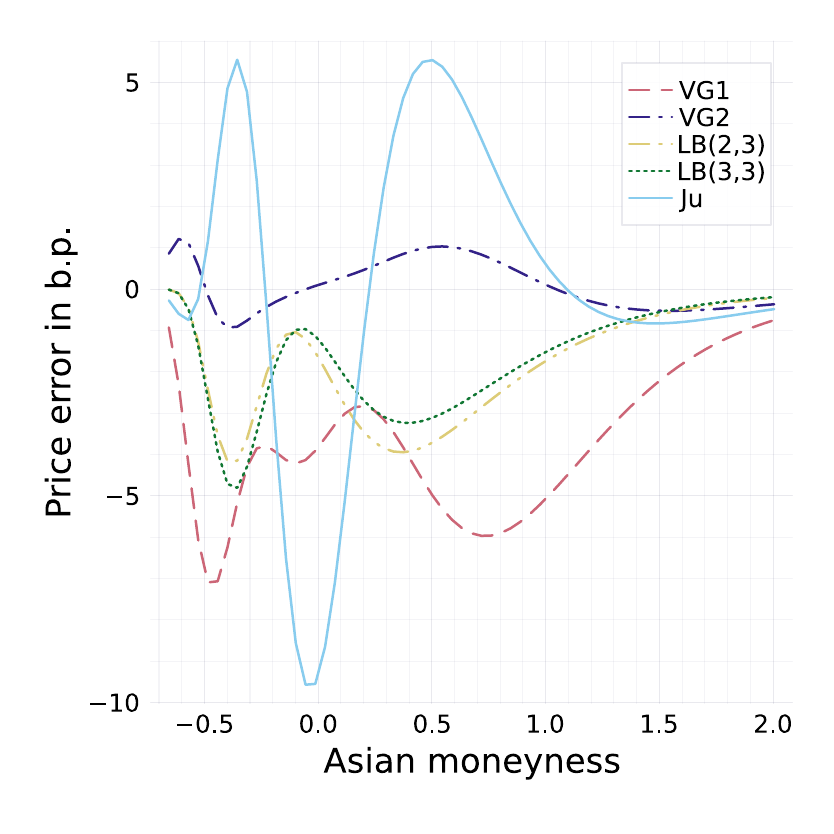}}
		\subfloat[][Second and third orders.]{\includegraphics[width=0.48\textwidth]{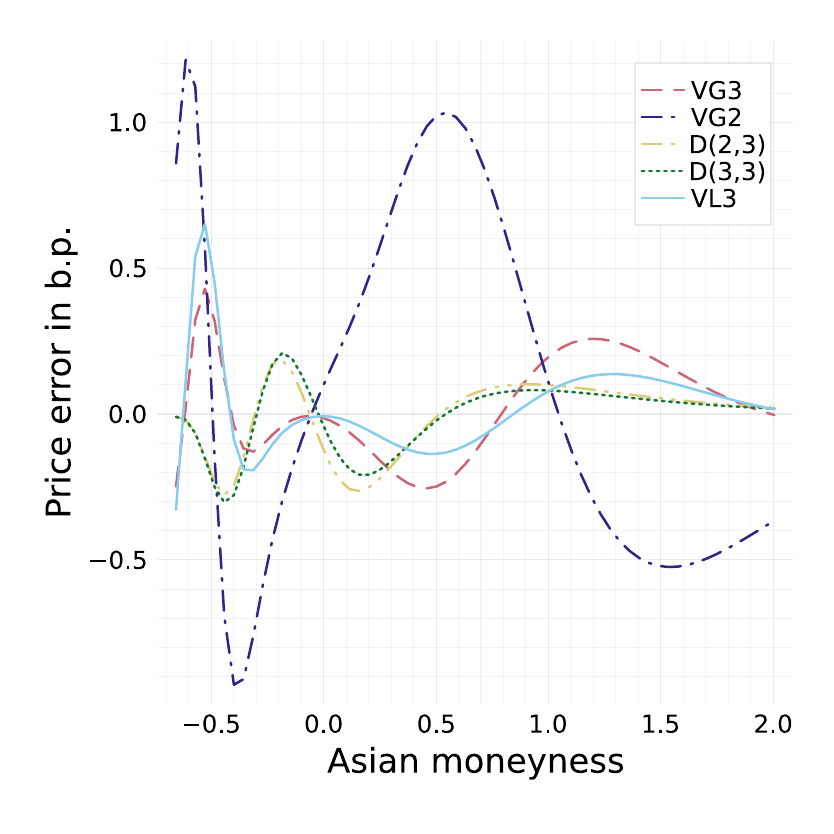}}
	}
	\caption{Error in the price of Asian options with observations at $t=0.1$ and $t=1.1$ year.}
	\label{fig:lord_yearly_2obs_second}
\end{figure}

We keep the same setting as the maturity 5 years in \cite{lord2005partially} but change the maturity to 1.1 year such that only two observations remain at $t=0.1$ and $t=T=1.1$. Figure \ref{fig:lord_yearly_2obs_second} plots the error for a wide range of strikes.

The Ju approximation leads to a maximum error of 10 b.p., twice as large as the Deelstra lower bound approximations. The first order expansion is slightly less accurate than the Deelstra lower bound and the second order expansion is significantly more accurate (with an maximum error of around 1 b.p.). The third order expansion is nearly as accurate as the Deelstra conditioned moment matching approximations.




\section{Numerical examples for basket options}
\subsection{Accuracy on basket options}
Krekel \cite{krekel2004analysis} tests various basket approximations in depth. We reproduce and apply some of the more significant tests to our new expansions.
\begin{table}[ht]
	\caption{Prices of vanilla basket options of maturity $T=5$ years, with homogeneous volatilities $\sigma_i=40\%$, varying the correlation $\rho_{i,j}$ for $i\neq j$. \label{tbl:krekeltable1}}
	\centering{
		\begin{tabular}{@{}l r r r r r r r r r@{}}\toprule
			$\rho$ & MC & Ju & MM3 & LB & Deelstra & VG1 & VG2 & VG3 &  VL3 \\ \midrule
0.10 & 21.692 & 21.766 & 21.619 &  20.124 & 21.609 & 20.124 & 22.224 & 21.440 & 21.612\\
0.30 & 25.029 & 25.052 & 24.994 &  24.209 & 24.987 & 24.209 & 25.212 & 24.961 & 24.985\\
0.50 & 28.007 & 28.013 & 27.995 &  27.633 & 27.990 & 27.633 & 28.059 & 27.994 & 27.996\\
0.70 & 30.743 & 30.743 & 30.740 &  30.620 & 30.738 & 30.620 & 30.752 & 30.741 & 30.742\\
0.80 & 32.041 & 32.041 & 32.041 &  31.989 & 32.040 & 31.989 & 32.044 & 32.041 & 32.041\\
0.95 & 33.919 & 33.919 & 33.919 &  33.916 & 33.919 & 33.916 & 33.919 & 33.919 & 33.919\\
		\midrule
			RMSE & & 0.032 &  0.034&  0.741 & 0.039 & 0.741 & 0.231 & 0.107 & 0.038\\
			MAE & & 0.073 & 0.073&  1.568& 0.083 & 1.568 & 0.532 & 0.252 & 0.081\\
			\bottomrule
	\end{tabular}}
\end{table}
The first test consists in a relatively long term option of maturity 5 years, using the same volatility for each asset and the same cross correlation, but varying this correlation from 10\% to 95\% (Table \ref{tbl:krekeltable1}). We include the results of a three moments matching to the shifted lognormal distribution (named MM3).

This very simple test already puts into evidence that the expansions on the VG proxy are accurate for correlations larger than 50\%, the third order being the most accurate, but deteriorate as the correlation decreases to 10\%. The expansions on the VL proxy are accurate across the whole range of correlations. Under the settings of this test, the first order expansion is equivalent to the Deelstra lower bound approximation.

\begin{table}[ht]
	\caption{Prices of vanilla basket options of maturity $T=5$ years, with homogeneous volatilities $\sigma_i=40\%$ and correlation $\rho_{i,j}=50\%$, for $i\neq j$, varying the strike $K$. \label{tbl:krekeltable2}}
	\centering{
		\begin{tabular}{l r r r r r r r r r}\toprule
			$K$ & MC & Ju & MM3 & LB & Deelstra & VG1 & VG2 & VG3 & VL3 \\ \midrule
		50.00 & 54.310 & 54.310 & 54.273 &  54.158 & 54.300 & 54.158 & 54.345 & 54.290 & 54.289\\
		60.00 & 47.481 & 47.482 & 47.446 &  47.270 & 47.468 & 47.270 & 47.524 & 47.459 & 47.459\\
		70.00 & 41.522 & 41.525 & 41.491 &  41.257 & 41.507 & 41.257 & 41.572 & 41.501 & 41.502\\
		80.00 & 36.351 & 36.355 & 36.327 &  36.041 & 36.335 & 36.041 & 36.404 & 36.332 & 36.334\\
		90.00 & 31.876 & 31.881 & 31.858 &  31.530 & 31.859 & 31.530 & 31.930 & 31.860 & 31.862\\
		100.00 & 28.007 & 28.013 & 27.995 &  27.633 & 27.990 & 27.633 & 28.059 & 27.994 & 27.996\\
		110.00 & 24.660 & 24.667 & 24.654 &  24.266 & 24.644 & 24.266 & 24.710 & 24.651 & 24.653\\
		120.00 & 21.762 & 21.769 & 21.761 &  21.356 & 21.746 & 21.356 & 21.808 & 21.756 & 21.758\\
		130.00 & 19.249 & 19.256 & 19.252 &  18.837 & 19.234 & 18.837 & 19.291 & 19.246 & 19.248\\
		140.00 & 17.065 & 17.072 & 17.071 &  16.652 & 17.052 & 16.652 & 17.102 & 17.065 & 17.066\\
		150.00 & 15.164 & 15.171 & 15.173 &  14.753 & 15.151 & 14.753 & 15.196 & 15.165 & 15.167\\	\midrule
			RMSE & & 0.005 & 0.021 & 0.347 & 0.015 & 0.347 & 0.045 & 0.014 & 0.013\\
			MAE & & 0.007 & 0.036 & 0.413 & 0.017 & 0.413 & 0.053 & 0.022 & 0.021\\
			\bottomrule
	\end{tabular}}
\end{table}
With a fixed correlation of $50\%$, the third order expansion has the same accuracy as the refined Deelstra approximation across a range of strikes (Table \ref{tbl:krekeltable2}). The Ju approximation is the most accurate on this example where the volatilities are, again, homogeneous.

Next we consider at-the-money options, with a constant correlation at 50\% but vary the homogeneous volatility from 5\% to 100\%. In practice,  volatilities larger than 50\% are exceptional for long term options. This test however puts in evidence the limits of applicability of some of the approximations.
The expansions become inaccurate for large volatility $\sigma_i > 50\%$ for a maturity of 5 years (Table \ref{tbl:krekeltable4}). The third order expansion clearly diverges as the volatility is increased.  This is true for the two proxies, the VL proxy being more accurate in general for $\sigma_i \leq 50\%$ but worse for larger volatilities.

Inhomogeneous volatilities (Table \ref{tbl:krekeltable5}) decrease the accuracy of the expansions slightly, but not significantly so, in contrast to the Ju approximation. This explains why the Ju approximation is not so good for Asian options. 
\begin{table}[ht]
	\caption{Prices of vanilla basket options of maturity $T=5$ years, with inhomogeneous volatilities $\sigma_2=\sigma_3=\sigma_4=\sigma$ but $\sigma_1=100\%$ and correlation $\rho_{i,j}=50\%$, for $i\neq j$, varying the volatility $\sigma$.\label{tbl:krekeltable5}}
	\centering{
		\begin{tabular}{@{}l r r r r r r r r r @{}}\toprule
			$\sigma$ & MC & Ju & MM3 & LB & Deelstra & VG1 & VG2 & VG3 & VL3\\ \midrule
			0.05 & 19.403 & 35.591 & 18.513 &  19.451 & 19.462 & 16.579 & 17.854 & 18.687 & 19.251\\
			0.10 & 21.092 & 36.189 & 18.646 &  20.838 & 20.998 & 18.822 & 19.934 & 20.542 & 20.836\\
			0.15 & 23.001 & 36.925 & 18.812 &  22.602 & 23.042 & 21.263 & 22.286 & 22.751 & 22.757\\
			0.20 & 25.304 & 37.801 & 19.016 &  24.694 & 25.400 & 23.836 & 24.823 & 25.209 & 24.987\\
			0.30 & 30.553 & 39.967 & 19.567 &  29.518 & 30.572 & 29.186 & 30.225 & 30.541 & 30.164\\
			0.40 & 36.023 & 42.664 & 20.379 &  34.722 & 35.980 & 34.601 & 35.841 & 36.031 & 35.806\\
			0.50 & 41.493 & 45.836 & 21.608 &  39.959 & 41.411 & 39.920 & 41.538 & 41.270 & 41.283\\
			0.60 & 46.733 & 49.393 & 23.566 &  45.047 & 46.746 & 45.036 & 47.264 & 45.719 & 45.907\\
			0.70 & 51.961 & 53.214 & 26.925 &  49.880 & 51.901 & 49.878 & 52.998 & 48.465 & 48.679\\
			0.80 & 56.792 & 57.171 & 33.186 &  54.394 & 56.802 & 54.394 & 58.733 & 47.745 & 47.711\\
			1.00 & 65.354 & 64.932 & 53.650 &  62.324 & 65.627 & 62.324 & 70.201 & 15.447 & 9.938\\

			\midrule
			RMSE & & 9.527 & 15.633 & 1.610 & 0.079 & 2.081 & 1.726 & 15.351 & 16.964\\
			MAE && 16.139 & 25.006 & 3.086 & 0.217 & 3.086 & 4.791 & 49.963 & 55.416\\
			\bottomrule
	\end{tabular}}
\end{table}
\subsection{Asian basket options on realistic market data}
\begin{figure}[h!]
	\centering{
		\subfloat[][6 months.]{\includegraphics[width=0.33\textwidth]{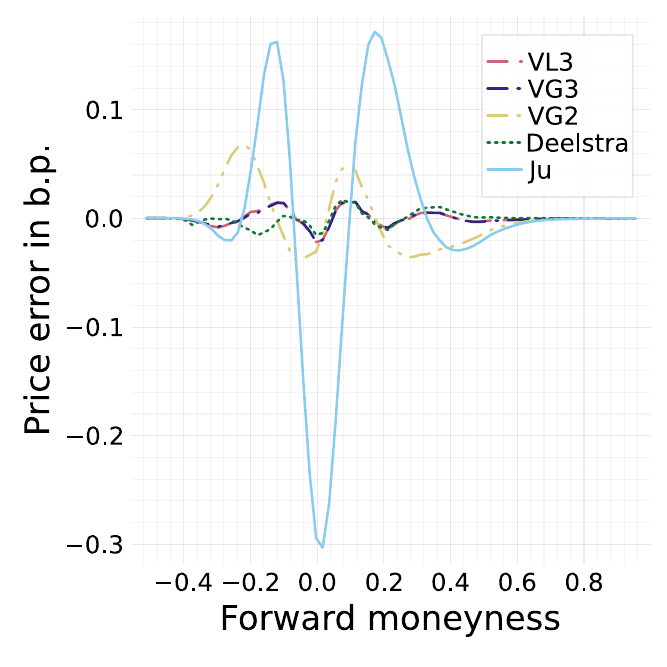}}
		\subfloat[][1 year.]{\includegraphics[width=0.33\textwidth]{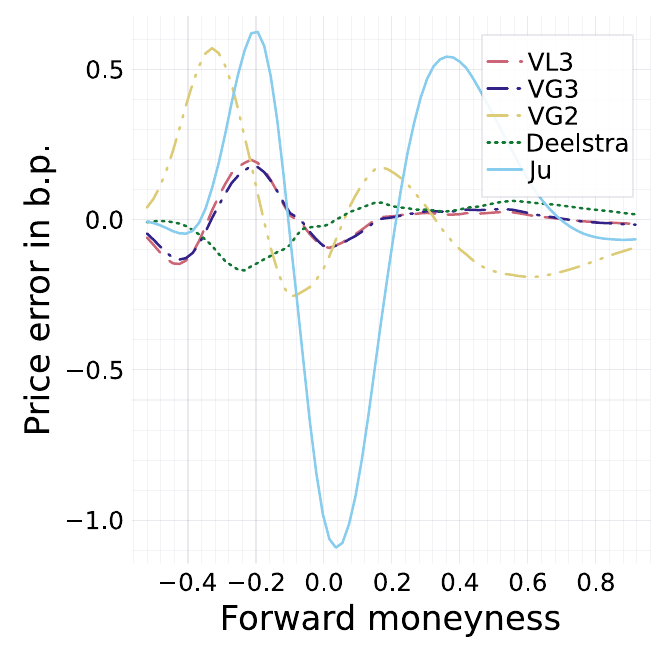}}
		\subfloat[][5 years.]{\includegraphics[width=0.33\textwidth]{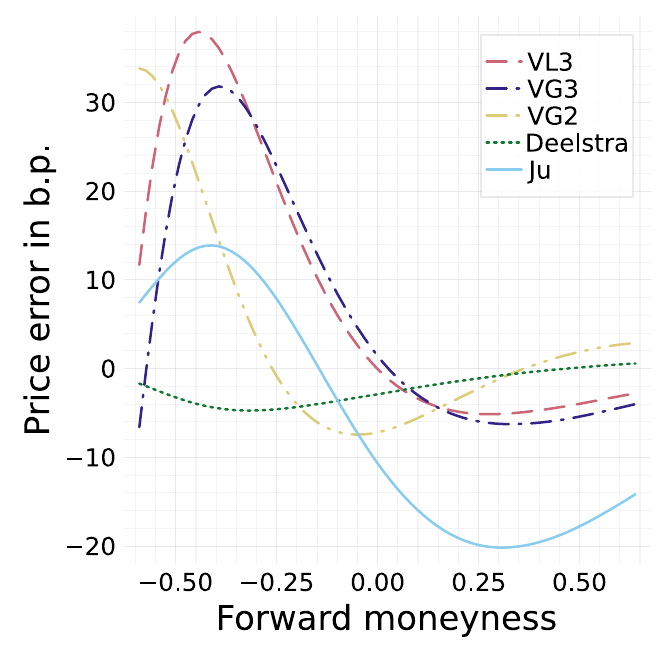}}
	}
	\caption{Error in the price of Asian basket options across a range of strikes, for three different maturities.}
	\label{fig:deelstra_asian_basket}
\end{figure}

Previously, we looked at the accuracy using simple, manufactured market data in order to assess the loss of accuracy when a single parameter was changing. In this section, we consider  Asian basket options on the more realistic market data of Beisser \cite{deelstra2010moment}. The basket consists in 5 stocks, with different forwards, volatilities and correlations. The averaging is monthly over the last 5 months. Figure \ref{fig:deelstra_asian_basket} shows the error in basis points \begin{equation*}\frac{V_{\textsf{approx}} -V_{\textsf{MC}}}{\sum_{j=1}^m \mu_j S_j(0)}\times 10000\,,\end{equation*} for three different maturities across a range of forward moneyness \begin{equation*}\frac{K}{\sum_{i=1}^n\sum_{j=1^m} w_i \mu_j F_j(0,t_i)}-1\,.\end{equation*}

Although not displayed, the first order expansion is found to have nearly the same error profile as the Deelstra lower bound approximation.
The higher order expansions are accurate for the 6 months and 1 year maturity, but degrade with the 5 years maturity, where the Deelstra conditioned moment matching leads to the smallest error. This is not too surprising in light of the previous tests:  the volatility is kept the same for the long term option and for the short term option, so the total variance is significantly larger in the long term case, which degrades the expansions accuracy. If we scale the volatility by the square root of the time to maturity, the error of the 5 years option price becomes very similar to the one  of  the 1 year option.

\section{Numerical examples for vanilla options with cash dividends}
Healy \cite{healy2021pricing} shows that the problem of pricing vanilla option under the piecewise constant lognormal model, where the underlying asset jumps from the cash dividend amount at the ex-dividend date, can be reduced to a vanilla basket option pricing problem under the Black-Scholes model. We may thus use our new expansions for basket options to price vanilla options with cash dividends and assess their accuracy against the stochastic expansions of Etor\'e, Gobet and Le Floc'h \cite{etore2012stochastic,lefloch2015more} and other discrete dividends approximations.

In Table \ref{tbl:vellekoop}, we reproduce the example of a 7-year option, with varying yearly dividend amounts, first presented in \cite{vellekoop2006efficient} and in \cite{etore2012stochastic}. 
\begin{table}[h]
	\centering{
		\caption{Price of a call option of maturity 7 years, on an asset with spot price $S=100$ paying yearly dividends of respective amounts 6, 6.5, 7, 7.5 8, 8, 8, starting at $t_1=0.9$, with $r=6\%$ and $\sigma=25\%$. \label{tbl:vellekoop}}
		\begin{tabular}{l c c c c c c c}\toprule
			Strike & FDM &  VL3 & VL2  & LL3 &EG3 & GS & Deelstra\\\midrule
			70 &  27.21395 &  27.21392 &27.21367&  27.21769 &27.20498  & 27.26457 & 27.21395 \\
			100 & 19.48229 & 19.48226 & 19.48181 & 19.48478 &19.47843 & 19.51519&19.48245 \\
			130 & 14.13026 &  14.13023&14.12969 &  14.13139 &14.12933 &14.15233&  14.13046 \\ \midrule
			MRE & N/A &   2E-6&4E-5 & 1E-4 & 3E-4 & 2E-3 & 1E-5\\
			\bottomrule
	\end{tabular}}
\end{table}
\begin{figure}[h]
	\centering{\includegraphics[width=\textwidth]{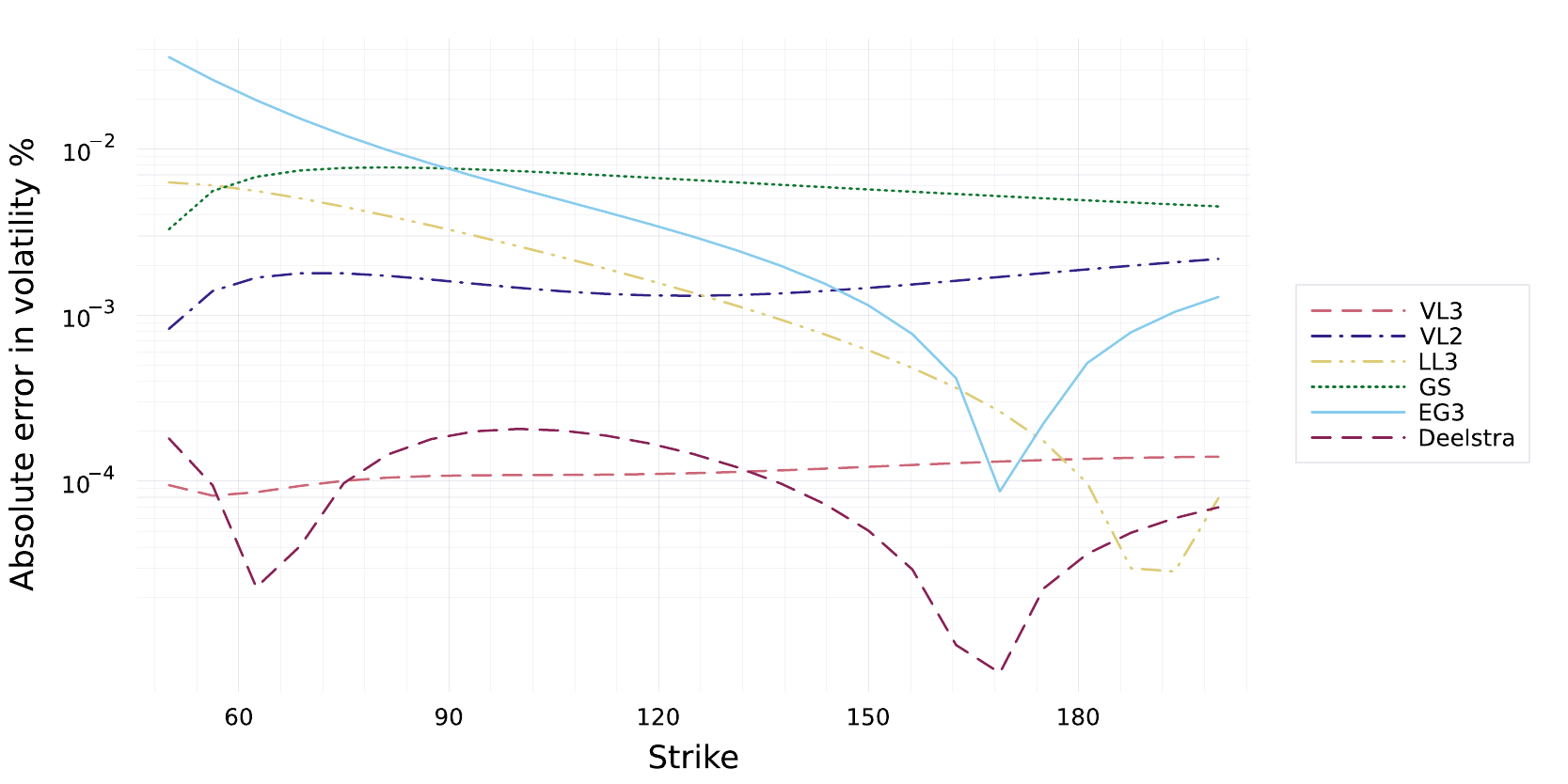}}
	\caption{Error in the price of vanilla options of maturity 10 years with semi-annual dividends for a range of strikes.}
	\label{fig:gs_pln_20}
\end{figure}
The reference, named "FDM" is the price obtained by the TR-BDF2 finite difference method using 10,000 space steps and 3,650 time-steps and is accurate to the fifth decimal. We also give the values of  the third order expansion from Etor\'e and Gobet \cite{etore2012stochastic} named "EG3", the third order expansion from Le Floc'h \cite{lefloch2015more} named "LL3" and the approximation from Gocsei and Sahel \cite{sahel2011matching} named "GS".

Figure \ref{fig:gs_pln_20} reproduces the example of a call option of maturity 10 years, on a stock of price $S=100$, paying dividends of amount $2$ on a semi-annual schedule, with the first dividend starting one day from the valuation date at $t_1 = 1/365$ presented in \cite{sahel2011matching}. The interest rate is taken to be $r=3\%$ and the volatility $\sigma=25\%$. 

On those two examples, the third order expansion on the Levy proxy is the most precise, and the second order expansion is more accurate than the third order expansions of Etor\'e, Gobet and Le Floc'h.

\section{Conclusion}
For Asian options, the second order expansion on the Vorst Geometric proxy is already very accurate, better than the third order formula of Ju or the Deelstra lower bound approximations. The third order expansion is comparable in accuracy to the more sophisticated Deelstra conditioning with moment matching. The stochastic expansions have the advantage of requiring no numerical solver and no numerical integration; they thus allow for a straightforward stable calculation of the greeks, and facilitate a robust implementation. The Vorst Levy proxy leads to a slightly  worse accuracy of the prices.

The expansions do not fare as well on basket options. In particular for very large (not so realistic) volatilities, the expansions misbehave, especially the third order expansion. The expansions on the Vorst Geometric proxy are also not so accurate under mild correlations. The Vorst Levy proxy remediates this latter issue. In the context of commodity options, the assets are typically futures on the same commodity but of different maturities, and have a correlation close to one \cite{clark2014commodity}. The payoff also often involves the arithmetic average of those futures prices: the contract is effectively an Asian basket option. Given such a correlation structure, the expansions would behave very well on those options.

When applied to price vanilla European options with cash dividends under the piecewise-lognormal model (by reducing the problem to a vanilla basket option pricing problem), the expansions are found to be the most precise against previously published approximations tailored to the specific case of cash dividends.

Further research may apply similar expansions to Asian spread options, using  a double geometric approximation as proxy (see also \cite{castellacci2003asian}). Other possible extensions would be for Asian options under discrete dividends assumptions, Asian options under stochastic volatility models with a reasonably fast pricing of European options such as Sch\"obel-Zhu.



\bibliographystyle{siamplain}
\bibliography{asian_expansion}

\appendix
\section{Derivatives of the Black-76 formula towards the strike}\label{sec:appendix_black_der}
The following formulae are used to compute the expansions in practice:
\begin{align}
	\frac{\partial}{\partial K} \Black(F, K, v, T) &= -\eta B(0,T)  \Phi(\eta d_2)\,,\\
	\frac{\partial^2}{\partial K^2} \Black(F, K, v, T) &= \frac{B(0,T)}{K\sqrt{v}}  \phi(d_2)\,,\\
	\frac{\partial^3}{\partial K^3} \Black(F, K, v, T) &= \frac{ B(0,T)}{K^2\sqrt{v}}  \left( \frac{\phi( d_2)}{\sqrt{v}}-1\right)\,,
\end{align}
where $\Phi$, $\phi$ are the cumulative normal distribution function and the normal probability density function.
 
\section{Additional tables} 
\begin{table}[ht]
	\caption{Prices of vanilla basket options of maturity $T=5$ years, with homogeneous volatilities $\sigma_i=\sigma\%$ and correlation $\rho_{i,j}=50\%$, for $i\neq j$, varying the volatility $\sigma$.\label{tbl:krekeltable4}}
	\centering{
		\begin{tabular}{@{}l r r r r r r r r r@{}}\toprule
			$\sigma$ & MC & Ju & MM3 & LB & Deelstra & VG1 & VG2 & VG3 & VL3 \\ \midrule
0.05 & 3.526 & 3.526 & 3.526 &  3.525 & 3.526 & 3.525 & 3.526 & 3.526& 3.526\\
0.10 & 7.050 & 7.050 & 7.050 &  7.043 & 7.050 & 7.043 & 7.050 & 7.050& 7.050\\
0.15 & 10.570 & 10.570 & 10.570 &  10.548 & 10.569 & 10.548 & 10.570 & 10.570& 10.570\\
0.20 & 14.083 & 14.083 & 14.083 &  14.032 & 14.082 & 14.032 & 14.085 & 14.083& 14.083\\
0.30 & 21.078 & 21.080 & 21.079 &  20.912 & 21.072 & 20.912 & 21.091 & 21.078& 21.078\\
0.40 & 28.009 & 28.013 & 27.995 &  27.633 & 27.990 & 27.633 & 28.059 & 27.994& 27.996\\
0.50 & 34.826 & 34.843 & 34.730 &  34.147 & 34.796 & 34.147 & 34.986 & 34.737& 34.750\\
0.60 & 41.488 & 41.519 & 41.044 &  40.412 & 41.450 & 40.412 & 41.881 & 41.070& 41.119\\
0.70 & 47.940 & 47.967 & 46.471 &  46.390 & 47.907 & 46.390 & 48.768 & 46.363& 46.502\\
0.80 & 54.128 & 54.095 & 50.433 &  52.050 & 54.120 & 52.050 & 55.705 & 48.888& 49.139\\
1.00 & 65.354 & 64.932 & 53.650 &  62.324 & 65.627 & 62.324 & 70.201 & 15.447& 9.938\\			\midrule
			RMSE && 0.145 & 3.745 & 1.281 & 0.069 & 1.281 & 1.547 & 15.155 & 16.782\\
			MAE && 0.478 & 11.760 & 3.086 & 0.217 & 3.086 & 4.791 & 49.963 & 55.416\\
			\bottomrule
	\end{tabular}}
\end{table}

\end{document}